\documentclass[rmp,11pt]{revtex4}
\usepackage{graphicx}
\begin{document}
\vspace*{2cm}
\title{Wormlike chain or tense string? A question of resolution}
\author{J.-B. Fournier}
\email{jbf@turner.pct.espci.fr}
\affiliation{Laboratoire de Physico-Chimie Th\'eorique, ESPCI,
10 rue Vauquelin, F-75231 Paris cedex 05, France}
\date{\today}
\begin{abstract}
It is shown that a wormlike chain, i.e., a filament with a fixed
contour-length~$\bar S$ and a bending elasticity~$\kappa$, attached to a
frame of length~$L$, can be described---at low resolutions---by the same
type of elastic free-energy as a tense string. The corresponding tension
is calculated as a function of temperature, $L$, $\kappa$ and $\bar S$.
\end{abstract}
\maketitle

Elastic free-energies are used to describe the distorted states and
fluctuations of soft-matter~\cite{chaikin_lubensky_book}. The latter
includes liquid-crystals~\cite{deGennes_book}, surfactant membranes or
interfaces~\cite{israelachvili_book}, and
polymers~\cite{doi_edwards_book}.  In this paper, we investigate how the
elastic free-energy functional of a filament with a fixed contour-length
and a bending rigidity (wormlike chain) depends on the {\em spatial
resolution\/} at which it is described. We perform a finite
renormalization group (RG) iteration in order to determine the
free-energy associated with the length of the filament that is hidden at
low-resolution. A similar study, focused on surfactant membranes, was
recently published in a short paper~\cite{fournier01}. 

The filaments considered here are known as \textit{wormlike
chains}~\cite{kratky49}, or \textit{semi-flexible
polymers}~\cite{landau_lifshitz_phystat_book,doi_edwards_book}. At
temperature~$T$, a filament with a bending rigidity~$\kappa$ has a
\textit{persistence length}
$L_p=\kappa/k_\mathrm{B}T$~\cite{landau_lifshitz_phystat_book}:
for contour-lengths $\bar S\ll L_p$ it resembles a flexible rod, while
for $\bar S\gg L_p$ it resembles a wandering polymer.  There are various
natural realizations of semi-flexible polymers: actin cytoskeletal
filaments, $L_p\simeq4-17\,\mu\textrm{m}$~\cite{kas94}, microtubules,
$L_p\simeq6\,\textrm{mm}$~\cite{gittes93}, and DNA,
$L_p\simeq50-350\,\textrm{nm}$~\cite{hagerman88,marko95}. Here
we shall consider fluctuating filaments with $\bar S\gg L_p$, however
they will be \textit{stretched} by attachment to a frame of width
$L\alt \bar S$. 

The paper is organized as follows: in Sec.~\ref{general} we recall how a
\textit{coarse-graining} procedure allows to define different
free-energies associated with different (spatial) resolutions of the
same system; in Sec.~\ref{high} we describe the high-resolution
free-energy of a wormlike chain; in Sec.~\ref{low} we calculate the
corresponding \textit{low-resolution} free-energy, and we show that it
coincides with that of a tense string; in Sec.~\ref{conclusion} we
summarize and discuss our results.

\section{Constructing elastic free-energies}
\label{general}

The free-energy of a system in thermodynamic equilibrium with a heath
bath at temperature~$T$ is defined by 
\begin{equation}
\label{FdeS}
F[{\cal S}]=-k_{\rm B}T\ln\left(
\sum_{i\in{\cal S}}e^{-\beta{\cal H}_i}
\right),
\end{equation}
where $\beta=1/k_{\rm B}T$ is the inverse temperature and $i$ labels the
microscopic states (microstates) of the system, ${\cal H}_i$ being the
energy, or Hamiltonian of state~$i$.  The above sum $\sum_{i\in{\cal
S}}\exp(-\beta{\cal H}_i)$, also called restricted partition function
$Z_\mathcal{S}$, runs over different subsets~$\cal S$ of the phase
space~$\Gamma$ of all the microstates of the system. Each subset~$\cal
S$ defines a {\em macrostate}: it is a collection of microstates that we
wish to consider macroscopically as a whole.  $F[{\cal S}]$ is the
free-energy of the macrostate~$\cal S$; it is either a functional or a
function depending on the explicit form of~$\cal S$. For instance, in
the standard thermodynamics of a fluid, the macrostates~$\cal S$ are
defined by specifying the number of particles~$N$ and the volume~$V$ of
the fluid. Hence, $F$ is a function of the variables $T,V,N$.

It is most convenient to divide the phase space~$\Gamma$ into a
collection of {\em disjoint} macrostates~$\cal S$.
Then, since we know from statistical mechanics that the probability for
the system to be in an arbitrary microstate~$i$ is given by
\begin{equation}
P_i=\frac{1}{Z_\Gamma}
e^{-\beta{\cal H}_i},
\end{equation}
where $Z_\Gamma=\sum_{i\in\Gamma}\exp(-\beta{\cal H}_i)$,
the quantity
\begin{equation}
P[\mathcal{S}]=\frac{1}{Z_\Gamma}e^{-\beta F[{\cal S}]}
\end{equation}
is the probability to find the system in the macrostate~$\cal S$.
Indeed, from Eq.\,(\ref{FdeS}) we have
$Z_\Gamma^{-1}\exp(-\beta F[{\cal S}])=
\sum_{i\in{\cal S}}Z_\Gamma^{-1}\exp(-\beta{\cal H}_i)$.
In other words, $F[{\cal S}]$ plays the role of a Hamiltonian for the
macrostates~$\cal S$. If the macrostates are well-chosen and large
enough, the system will fluctuate little around the macrostate~${\cal
S}^\star$ that minimizes $F[{\cal S}]$, i.e., around the most probable
macrostate. The latter determines then the macroscopic ``equilibrium
state'' of the system.

In general, for a soft-matter system, one chooses for the
macrostates~$\cal S$ the profiles of the {\em order-parameter} field:
the latter is defined as the local average of some interesting material
property, e.g., for a nematic\footnote{a fluid with a long-range
non-polar orientational order of its constituent molecules.},
$Q_{ij}({\bf r})=\langle m_i({\bf x})m_j({\bf
x})-\frac{1}{3}\delta_{ij}\rangle$, where ${\bf m}({\bf x})$ is the
orientation of the molecule situated at point~$\bf x$, the average
running over a small volume centered on $\mathbf{r}$.  Note that the
size $\Lambda_0^{-1}$ of this volume may be chosen at will and sets a
cutoff $\Lambda_0$ for the order-parameter field. In the case of a
filament, one may simply choose the positions $\mathbf{r}(s)$ of
its constituting elements.

It is essential to note that we have a complete freedom to choose the
``resolution'' at which the phase space~$\Gamma$ is coarse-grained into
the macrostates~$\cal S$. For example, on the one hand, we may be
interested in the {\em microscopic} fluctuations of a bilayer membrane,
as resolved at the scale of the membrane thickness ($\simeq40$\AA). We
would then define the free-energy $F[{\cal S}]$, as in Eq.\,(\ref{FdeS}),
by summing over all the microstates~$i$ compatible with a given
shape~$\cal S$ of the bilayer. Specifying one of these microstates would
require specifying all the coordinates, velocities and conformations of
all the lipids within the bilayer. On the other hand, we may be
micro-manipulating a membrane under an optical microscope. In this case,
we would only be interested in the Fourier components of the membrane
shape with wavevector~$q$ less than the typical wavevector of the
visible light. We would then construct a free-energy functional $\bar
F[\Sigma]$ by summing over all the microstates compatible with the
poorly resolved shape~$\Sigma$ of the membrane~\cite{fournier01}. This
would give us a tool more adapted to the experiment, since
$\exp(-\beta\bar F[\Sigma])$ is directly the probability
of observing $\Sigma$ (up to a normalization
constant).

The free energy $\bar F[\Sigma]=-k_{\rm
B}T\ln\sum_{i\in\Sigma}\exp(-\beta{\cal H}_i)$ associated with the
coarse-grained
states $\Sigma$ can be calculated also from
\begin{equation}
\label{FbardeSigmadirect}
\bar F[\Sigma]=-k_{\rm B}T\ln\left(
\sum_{{\cal S}\in\Sigma}e^{-\beta F[{\cal S}]}
\right),
\end{equation}
since
\begin{equation}
\label{cgcg}
\sum_{i\in\Sigma}e^{-\beta{\cal H}_i}
=\sum_{{\cal S}\in\Sigma}\sum_{i\in{\cal S}}e^{-\beta{\cal H}_i}
=\sum_{{\cal S}\in\Sigma}e^{-\beta F[{\cal S}]}.
\end{equation}
Equations (\ref{FbardeSigmadirect}) and~(\ref{cgcg}) define a
``coarse-graining'' procedure that can be carried-out for any partition
$\Sigma$ of the phase space that is obtained by grouping the elements of
a sub-partition $\cal S$. Note that the RG consists in applying this
procedure repeatedly while rescaling the lengths and the fields at each
step~\cite{goldenfeld_book}.

It is usually impossible to directly perform the microscopic sum
defining $\bar F[\Sigma]$. However, assuming on the grounds of
symmetry\footnote{One usually represents the macrostates $\mathcal{S}$
by a small deformation field $\phi$, and one writes $F$ as the integral
of a free-energy density expanded in power series of $\phi$ and its
gradients.} a simple form for $F[{\cal S}]$, it may be possible to
determine $\bar F[\Sigma]$ analytically from
Eq.~(\ref{FbardeSigmadirect}). The functional $\bar F[\Sigma]$ will in
general involve {\em renormalized\/} elastic
constants~\cite{chaikin_lubensky_book} depending both on the elastic
constants of $F[{\cal S}]$ and on temperature~$T$ (these new elastic
constants take into account the effects of the thermal fluctuation
occurring at shorter scales).  However, $\bar F[\Sigma]$ may have a
functional form fairly different from that of $F[{\cal S}]$, as we shall
see in the following. 

It should be noted that this procedure, although akin to that of the RG,
is different from the point of view of its scope.  The scope of the RG
is to identify the phases and critical points of a system by
coarse-graining/rescaling infinitely. Here, the scope is to construct a
free-energy functional {\em adapted\/} to the resolution at which the
system is actually observed, or described (i.e., a tool fit to determine
the shape and fluctuations of the system at this resolution).

\section{A wormlike chain at high resolution}
\label{high}

\begin{figure}
\includegraphics[width=.6\columnwidth]{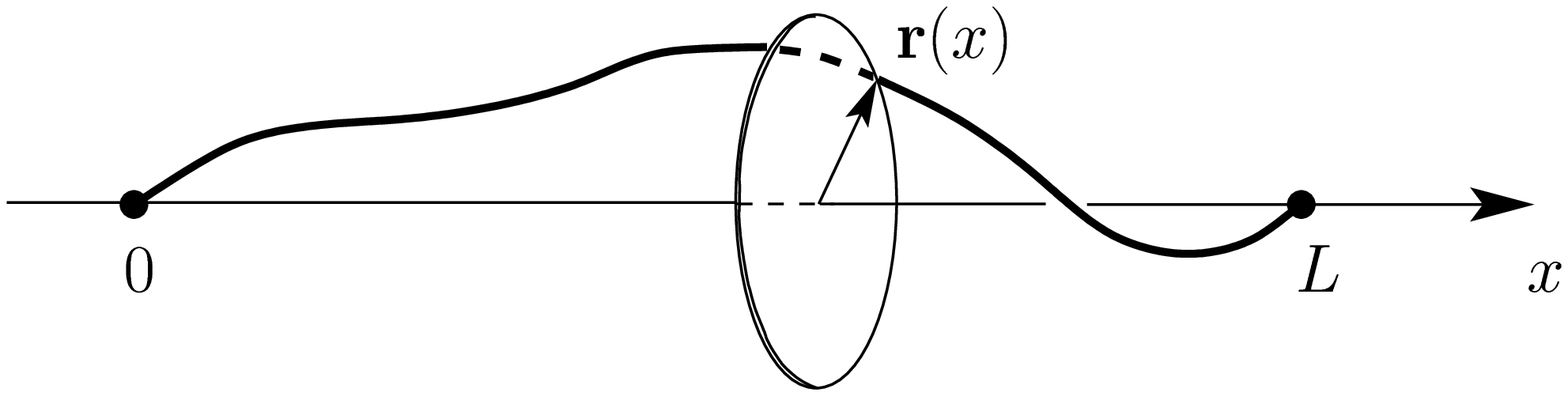}
\caption{Parameterization of the filament's shape. The disc indicates the
plane normal to ${\bf\hat x}$, in which ${\bf r}(x)$ lives. In practice,
we shall consider filaments that remain in the vicinity of the $x$-axis.}
\label{param}
\end{figure}

Let us consider a filament of fixed contour-length $\bar S$, attached at
its extremities to two fixed points, separated by a distance $L$ along
the $x$-axis (Fig.~\ref{param}). We choose as macrostates~$\cal S$ (see
Sec.\,\ref{general}) the {\em microscopic shape} of the filament,
measured at the resolution that corresponds to its thickness. We
restrict our attention to small deformations of the filament, hence we
parameterize its shape by a normal displacement field ${\bf r}(x)$, with
${\bf r}\cdot{\bf\hat x}=0$ and $x\in[0,L]$.  Each macrostate~$\cal S$
thus corresponds to a field ${\bf r}(x)$. Let us introduce its Fourier
transform ${\bf r}_q$: extending ${\bf r}(x)$ periodically in such a way
that ${\bf r}(x+L)={\bf r}(x)$ for each $x$, we may write
\begin{eqnarray}
\label{rFourier}
{\bf r}(x)&=&\frac{1}{\sqrt{L}}\sum_q {\bf r}_q\,e^{iqx},\\
{\bf r}_q&=&\frac{1}{\sqrt{L}}\int_0^L\!dx\,{\bf r}(x)e^{-iqx},
\end{eqnarray}
with $q=n\times2\pi/L$, where $n$ is a positive or negative integer.
Since the highest resolution for ${\bf r}(x)$ corresponds to the
filament's thickness $\simeq\!a$, we require
\begin{equation}
|q|<\Lambda_0\equiv\frac{2\pi}{a}.
\end{equation}

The filament is assumed to be made of a fixed number of incompressible
segments, hence its contour-length
\begin{equation}
\bar S=\int_0^L\!dx\,\frac{ds}{dx}\simeq
L+\int_0^L\!dx\,\frac{1}{2}\left(\frac{d{\bf r}}{dx}\right)^2
\end{equation}
is fixed. Here we have used the exact expression for the elementary
arc-length $ds=dx\sqrt{1+(d{\bf r}/dx)^2}$ and we have expanded the
result up to second-order in ${\bf r}(x)$. In the following, assuming
small tranverse excursions of the filament, we shall systematically
expand all expressions in power-series of ${\bf r}(x)$, and retain only
second-order terms.

Using Eq.\,(\ref{rFourier}) and $\int_0^L\!dx\exp(iqx)=L\delta_{q,0}$,
where $\delta$ is the Kronecker symbol, we have
\begin{equation}
\bar S-L\simeq\sum_{|q|<\Lambda_0}\frac{1}{2}q^2
\left|{\bf r}_q\right|^2.
\end{equation}
We have used the property ${\bf r}_{-q}={{\bf r}_q}^\star$ that follows
from the fact that ${\bf r}(x)$ is real-valued.

We assume that the free-energy of the filament, obtained in principle
from the microscopic material properties through the procedure embodied
in Eq.\,(\ref{FdeS}), is simply proportional to the integral of the
square of the filament's curvature~\cite{kratky49,harris66}. More precisely,
\begin{eqnarray}
\label{Fcurv}
F[{\bf r}]&=&\int_0^{\bar S}\!ds\,\frac{1}{2}\kappa\left(\frac{d{\bf
t}}{ds}\right)^2\quad\quad
{\rm if}\,\int_0^L\!dx\,\frac{ds}{dx}=\bar S,\\
\label{Fcurvinf}
F[{\bf r}]&=&\infty\quad\quad\quad\quad\quad\quad\quad\quad
{\rm otherwise}.
\end{eqnarray}
Let us calculate the squared curvature up to second-order in $\bf r$.
From $d{\bf M}=dx\,{\bf\hat x}+d{\bf r}$, where $\bf M$ is a vector the
extremity of which runs along the filament, we deduce the tangent vector
${\bf t}=d{\bf M}/ds\simeq[1-\frac{1}{2}(d{\bf r}/dx)^2]{\bf\hat
x}+d{\bf r}/dx$. Hence, the curvature is $d{\bf t}/ds\simeq-[(d^2{\bf
r}/dx^2)\cdot(d{\bf r}/dx)]\,{\bf\hat x}+d^2{\bf r}/dx^2$ and $(d{\bf
t}/ds)^2\simeq(d^2{\bf r}/dx^2)^2$. The functional in Eq.\,(\ref{Fcurv})
becomes, up to second order in $\bf r$,
\begin{equation}
\label{nrj}
F[{\bf r}]\simeq\int_0^L\!dx\,\frac{1}{2}\kappa
\left(\frac{d^2{\bf r}}{dx^2}\right)^2=
\sum_{|q|<\Lambda_0}\frac{1}{2}\kappa q^4\left|{\bf r}_q\right|^2.
\end{equation}

\section{The same wormlike chain at low resolution}
\label{low}

Let us now assume that we are not interested in the Fourier components
of the filament's shape in the range $\Lambda<|q|<\Lambda_0$ (with
$\Lambda>0$).  For instance, we may be observing the filament with an
experimental setup of resolution $2\pi/\Lambda$. The coarse-grained
macrostates~$\Sigma$ of interest (see Sec.\,\ref{general}) correspond then
in real space to
\begin{equation}
\mathbf{R}({\bf x})=\sum_{|q|<\Lambda}
\mathbf{r}_q\,e^{iqx}.
\end{equation}
The field ${\bf R}(x)$, i.e., the poorly resolved shape of the wormlike
chain, has a cutoff $\Lambda$ lower than that of ${\bf r}(x)$. We are
interested in the free-energy $\bar F[{\bf R}]$ defined as in
Eq.~(\ref{FbardeSigmadirect}) and {\em not} in $F[{\bf r}]$. Indeed,
$\exp\{-\beta\bar F[{\bf R}]\}$ is, up to a multiplicative constant, the
probability for the occurrence of a given low-resolution shape ${\bf
R}(x)$---with any possible wrinkle in the either inaccessible or
uninteresting range $\Lambda<|q|<\Lambda_0$; on the contrary,
$\exp\{-\beta F[{\bf R}]\}$ would be the probability for the occurrence
of a given low-resolution shape ${\bf R}(x)$ having no high-wavevector
wrinkles at all (which is useless since it corresponds to a very
particular and unlikely event). Note that in a purely harmonic
system---which is not the case here because of the fixed contour-length
constraint---the probability weight of the wrinkles is actually
independent of the low-resolution shape ${\bf R}(x)$; in this case $\bar
F[{\bf R}]=F[{\bf R}]$.

Following the procedure embodied in Eq.\,(\ref{FbardeSigmadirect}), we
have
\begin{equation}
\label{Fbar}
\bar F[{\bf R}]=-k_{\rm B}T
\ln
\int\!\!\prod_{\Lambda<q<\Lambda_0}
\!\!d{\bf r}'_q\,d{\bf r}''_q\,\,
\delta\!\left(
L+\sum_{|q|<\Lambda_0}\frac{1}{2}q^2|{\bf r}_q|^2-\bar S
\right)
\exp\left(
-\beta\sum_{|q|<\Lambda_0}\frac{1}{2}\kappa q^4|{\bf r}_q|^2
\right).
\end{equation}
Here, ${\bf r}'_q$ and~${\bf r}''_q$ are the real and imaginary parts of
${\bf r}_q$, respectively. The integration variables [the wrinkles
hidden in ${\bf R}(x)$] span the range $\Lambda<q<\Lambda_0$ of {\em
positive} wavevectors only, because ${\bf r}_{-q}$ and ${\bf r}_q$ are
not independent (they are complex conjugates). In principle,
the above calculation is valid only for small values of ${\bf r}_q$,
however, thanks to the Gaussian character of the integrand, we take the
integrals as running over $]-\infty,\infty[$. The Dirac delta-function
takes into account the constraint on the filament's contour length [see
Eq.\,(\ref{Fcurvinf})].

Let us call $S$ the contour-length of the low-resolution shape ${\bf
R}(x)$, i.e., the {\em apparent} contour-length. The latter, which is
{\em no longer fixed\/}, is given by
\begin{equation}
S\simeq L+\sum_{|q|<\Lambda}\frac{1}{2}q^2|{\bf r}_q|^2.
\end{equation}
Using this relation and the integral representation
$\delta(x)=\int_{-\infty}^\infty\!d\lambda\exp(i\lambda x)$ of the
delta-function, we may rewrite Eq.\,(\ref{Fbar}) in the form
\begin{eqnarray}
\exp\left\{-\beta\bar F[{\bf R}]\right\}&=&
\int_{-\infty}^\infty\frac{d\lambda}{2\pi}
\int\!\!\prod_{\Lambda<q<\Lambda_0}
\!\!d{\bf r}'_q\,d{\bf r}''_q\times\nonumber\\*
&&\exp\left[
i\lambda\left(
S-\bar S+\sum_{\Lambda<|q|<\Lambda_0}\frac{1}{2}q^2|{\bf r}_q|^2
\right)
-\beta\sum_{|q|<\Lambda_0}\frac{1}{2}\kappa q^4|{\bf r}_q|^2
\right].
\end{eqnarray}
Since ${\bf r}_q={\bf R}_q$ for $|q|<\Lambda$, we have
\begin{eqnarray}
\exp\left\{-\beta\bar F[{\bf R}]\right\}&=&
\exp\left(
-\beta\sum_{|q|<\Lambda}\frac{1}{2}\kappa q^4|{\bf R}_q|^2
\right)
\int_{-\infty}^\infty\frac{d\lambda}{2\pi}\,
e^{i\lambda(S-\bar S)}\times\nonumber\\
&&\int\!\!\prod_{\Lambda<q<\Lambda_0}
\!\!d{\bf r}'_q\,d{\bf r}''_q
\exp\left\{
\frac{1}{2}\sum_{\Lambda<|q|<\Lambda_0}
\left(i\lambda q^2-\beta\kappa q^4\right)
\left[
\left({\bf r}'_q\right)^2
+\left({\bf r}''_q\right)^2
\right]
\right\}.
\end{eqnarray}
The rightmost integral, which is simply a product of separate Gaussian
integrals, is equal to
\begin{equation}
\label{Gaussianintegrals}
\left(
\prod_{\Lambda<q<\Lambda_0}\sqrt{\frac{\pi}
{\beta\kappa q^4-i\lambda q^2}}
\right)^4
=\exp\left[
2\sum_{\Lambda<q<\Lambda_0}\ln\left(
\frac{\pi}
{\beta\kappa q^4-i\lambda q^2}
\right)\right].
\end{equation}
For large values of $L$, we may replace the discrete sum by an integral.
Hence Eq.\,(\ref{Gaussianintegrals}) can be approximated by 
\begin{equation}
\exp\left[
-\frac{L}{\pi}
\int_\Lambda^{\Lambda_0}\!dq\ln\left(
\beta\kappa q^4-i\lambda q^2
\right)\right],
\end{equation}
up to a multiplicative constant independent of $\lambda$, which we
discard as it merely shifts $\bar F[{\bf R}]$ by a fixed amount
independent of ${\bf R}(x)$. We obtain
\begin{eqnarray}
\exp\left\{-\beta\bar F[{\bf R}]\right\}&=&
\exp\left(
-\beta\sum_{|q|<\Lambda}\frac{1}{2}\kappa q^4|{\bf R}_q|^2
\right)\times\nonumber\\
&&\int_{-\infty}^\infty d\lambda
\exp\left[
i\lambda(S-\bar S)
-\frac{L}{\pi}
\int_\Lambda^{\Lambda_0}\!dq\ln\left(
\beta\kappa q^4-i\lambda q^2
\right)\right],
\end{eqnarray}
yielding
\begin{eqnarray}
\label{FbaravecphideS}
\bar F[{\bf R}]&=&\int_0^L\!dx\,\frac{1}{2}\kappa
\left(
\frac{d^2{\bf R}}{dx^2}
\right)^2+\phi(S),\\
\label{phideSexact}
\phi(S)&=&-\frac{1}{\beta}
\ln\int_{-\infty}^\infty d\lambda
\exp\left[
i\lambda(S-\bar S)
-\frac{L}{\pi}
\int_\Lambda^{\Lambda_0}\!dq\ln\left(
\beta\kappa q^4-i\lambda q^2
\right)\right].
\end{eqnarray}
Thus the free-energy associated to ${\bf R}(x)$ contains a bending term
[as previously for $\mathrm{r}(x)$] plus a \textit{new} potential
$\phi(S)$ function of the contour-length of the low-resolution
filament's shape. Note that corrections to $\kappa$ would have been
generated if we had taken into account the non-harmonic terms in
Eq.~(\ref{nrj})~\cite{helfrich85,peliti85}. A correct treatment of these
terms, including measure factors~\cite{cai94} is outside the scope of
this paper.

In the thermodynamic limit ($L\to\infty$), the integral in $\phi(S)$ can
be approximated by its saddle-point value~\cite{walker_book}. This yields
\begin{equation}
\label{phideS1}
\phi(S)\simeq
-\frac{i\hat\lambda(S)}{\beta}(S-\bar S)
+\frac{L}{\pi\beta}\int_\Lambda^{\Lambda_0}\!dq\ln\left(
\beta\kappa q^4-i\hat\lambda(S)q^2
\right),
\end{equation}
with $\hat\lambda(S)$ the value of $\lambda$ that maximizes the
integrand in Eq.\,(\ref{phideSexact}); $\hat\lambda(S)$ is defined by
\begin{equation}
\label{phideS2}
\frac{\bar S-S}{L}=
\frac{1}{\pi}\int_\Lambda^{\Lambda_0}
\frac{dq}{\beta\kappa q^2-i\hat\lambda}.
\end{equation}
Note that $\hat\lambda(S)$ actually lies on the imaginary axis of the
complex plane. The Eqs.~(\ref{phideS1}) and~(\ref{phideS2}) define
$\phi(S)$ implicitly.

A quantity of interest is
\begin{equation}
\sigma(S)=\frac{d\phi(S)}{dS}.
\end{equation}
It corresponds to the {\em line tension} associated with the
low-resolution filament's shape ${\bf R}(x)$. Let us call
$\tilde\phi(S,\hat\lambda)$ the right-hand side of Eq.\,(\ref{phideS1}).
Then we have
\begin{equation}
\sigma(S)=\frac{d\tilde\phi(S,\hat\lambda(S))}{dS}
=\frac{\partial\tilde\phi}{\partial S}
+\frac{\partial\tilde\phi}{\partial\hat\lambda}
\frac{d\hat\lambda}{dS}.
\end{equation}
Since, by definition of $\hat\lambda$,
$\partial\tilde\phi/\partial\hat\lambda=0$, we obtain
\begin{equation}
\sigma(S)=-\frac{i\hat\lambda(S)}{\beta}.
\end{equation}
Thus, the central result is the following: the line tension $\sigma(S)$,
the primitive of which is the potential $\phi(S)$, is given by
\begin{equation}
\label{sigmadeS}
\pi\beta\kappa\frac{\bar S-S}{L}=
\int_\Lambda^{\Lambda_0}
\frac{dq}{q^2+\displaystyle\frac{\sigma(S)}{\kappa}}.
\end{equation}

\subsection{Analysis}

\begin{figure} 
\includegraphics[width=.6\columnwidth]{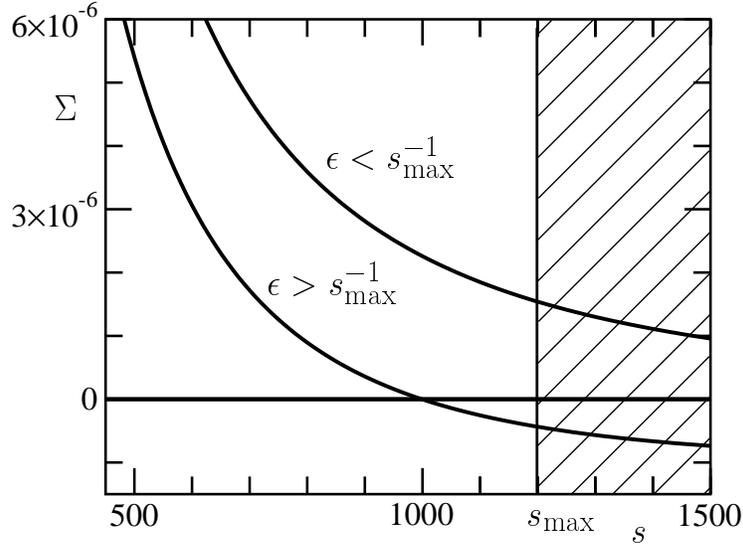}
\caption{Plot of the reduced effective wormlike chain's tension~$\Sigma$
against the reduced hidden length~$s$. The hatched region is forbidden.
The top curve corresponds to $\epsilon=10^{-4}$ and the lower one to
$\epsilon=10^{-3}$. For $\epsilon<s_{\rm max}^{-1}$ there is no way to
obtain a vanishing tension: the filament behaves as a tense ``string''.}
\label{tension} \end{figure}

Let us introduce the following dimensionless quantities:
\begin{eqnarray}
\label{s}
s&=&\pi\beta\kappa\Lambda_0\frac{\bar S-S}{L},\\
\epsilon&=&\frac{\Lambda}{\Lambda_0},\\
\Sigma&=&\frac{\sigma}{\kappa\Lambda_0^2}.
\end{eqnarray}
The parameter~$s$ measures the hidden length of the filament. Since
$L<S<\bar S$, we have
\begin{equation}
\label{smax}
0<s<s_{\rm max},\quad\quad
{\rm where}\quad
s_{\rm max}=\pi\beta\kappa\Lambda_0\frac{\bar S-L}{L}.
\end{equation}
The dimensionless number $\beta\kappa\Lambda_0$, appearing in
Eqs.~(\ref{s}) and~(\ref{smax}), is the ratio of the persistence length
of the filament, $L_p=\beta\kappa$, to its thickness
$\approx\!\Lambda_0^{-1}$. For a wormlike chain, we expect
\begin{equation}
\beta\kappa\Lambda_0\gg1.
\end{equation}
The parameter~$\epsilon$ is the ratio of the high-resolution scale (the
filament thickness) to the low-resolution scale $\Lambda^{-1}$.
Obviously, we shall consider the case
\begin{equation}
\epsilon\ll1.
\end{equation}
The parameter~$\Sigma$ (having now a different meaning than in
Sec.~\ref{general}) measures the filament's tension in units of
$\kappa\Lambda_0^2$. Note that for $\beta\kappa\Lambda_0\gg1$, the
quantity $\kappa\Lambda_0^2$ is much larger than $k_{\rm B}T\Lambda_0$,
which we would expect if the tension were of molecular origin.

In terms of these dimensionless quantities, the filament's tension,
$\Sigma(s)$, is defined by the equation:
\begin{equation}
\label{Sigmades}
s=\int_\epsilon^1\frac{dk}{k^2+\Sigma}.
\end{equation}
Let us now study the various regimes displayed by $\Sigma(s)$. For
$s\to0$, $\Sigma$ diverges as $(1-\epsilon)/s$, as can be easily seen by
neglecting $k^2$ with respect to $\Sigma$ in Eq.\,(\ref{Sigmades}).
$\Sigma$ decreases monotonically and vanishes at $s=s^\star(\epsilon)$
with (for $\epsilon\ll1$)
\begin{equation}
s^\star\simeq\frac{1}{\epsilon};
\end{equation}
it becomes then negative. When $\epsilon<s_{\rm max}^{-1}$, $s^\star$
lies beyond $s_{\rm max}$ and is thus unreachable (Fig.\,\ref{tension}).
Hence, as in the membrane case discussed by \textcite{fournier01}, there are
two regimes: for $\epsilon>s_{\rm max}^{-1}$, i.e., when the
resolution~$\Lambda$ is higher than $\Lambda_0/s_{\rm max}$, the
filament is in a {\em floppy} regime, while for $\epsilon<s_{\rm
max}^{-1}$ (low resolution), it is in a {\em tense} regime.

In the floppy regime, the low-resolution contour ${\bf R}(x)$ exhibits a
loose shape with an apparent contour-length fluctuating
in the vicinity of the value $S^\star$ (associated to $s^\star$) that
minimizes the potential $\phi(S)$. At $S=S^\star$, we have $\sigma=0$:
the tension effectively vanishes. This is reminescent of the case of
membranes~\cite{david91,seifert95}.

\subsection{Effectively a tense string}

\begin{figure}
\includegraphics[width=.6\columnwidth]{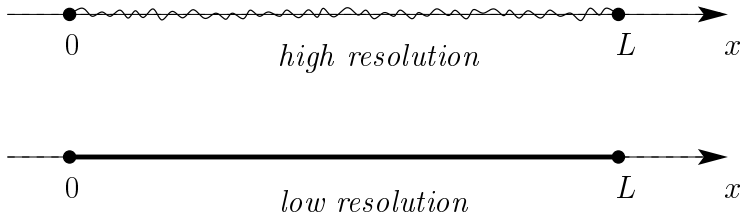}
\caption{The same wormlike chain viewed at high and low
resolutions. The elasticity describing the low-resolution shape and its
fluctuations (not drawn) is similar to that of a tense string.} 
\label{highetlow}
\end{figure}

In the tense regime, the apparent contour-length minimizing $\phi(S)$
corresponds to the lowest possible value of $\Sigma$, attained at
$s=s_{\rm max}$, which corresponds to the lowest possible value of $S$,
i.e., $L$. The low-resolution, apparent shape of the filament is thus
the {\em straight line} joining the two fixed extremities
(Fig.\,\ref{highetlow}). The filament, described at such low
resolutions, behaves therefore effectively as a tense string of tension
$\sigma_0=\sigma(L)$. As one pulls on the ``string'', the apparent
contour-length increases ($S>L$) and so does the tension.

Although Eq.~(\ref{Sigmades}) cannot be solved analytically for
$\Sigma$, approximate solutions can be obtained in the tense regime both
for medium and high tensions. Assuming $\epsilon\ll1$, we replace the
lowest bound of the integral in Eq.\,(\ref{Sigmades}) by zero. This
gives $s\simeq(1/\sqrt{\Sigma})\arctan(1/\sqrt{\Sigma})$, which yields
the following approximations: $s\simeq\pi/(2\sqrt{\Sigma})$ for
$0<\Sigma\ll1$ (medium tensions), and $s\simeq1/\Sigma$ for $\Sigma\gg1$
(high tensions).

For real wormlike chains and in the high tension case, we should
actually take into account the finite stretching elasticity of the
filament: this would modify our results. We shall therefore concentrate
on the medium tension states of the tense regime. The relation
$s\simeq\pi/(2\sqrt{\Sigma})$ yields
\begin{equation}
\label{sigmalow}
\sigma\simeq\frac{1}{4\beta^2\kappa}\left(\frac{L}{\bar S-S}\right)^2.
\end{equation}
This represents the free-energy we have to pay in order to increase the
apparent contour-length of the filament by a unit amount. It is worth
noticing the intrinsic character of this expression: it is independent
both of the microscopic cutoff~$\Lambda_0$ and of the
resolution~$\Lambda$ at which the filament is observed. For
Eq.\,(\ref{sigmalow}) to hold, it is necessary that
\begin{equation}
\label{rangevalid}
\Lambda\ll\Lambda_c=\frac{1}{\pi\beta\kappa}\frac{L}{\bar S-L},
\end{equation}
in order to be well into the tense regime ($\epsilon\ll s_{\rm
max}^{-1}$).

\section{Conclusion}
\label{conclusion}

We have shown that at spatial resolutions~$\Lambda$ much less than the
wavevector~$\Lambda_c$ defined above, a wormlike chain, i.e., a filament
with a bending elasticity~$\kappa$ and a fixed contour-length~$\bar S$,
behaves as a {\em tense string} when it is attached at its extremities
to a frame of length~$L$. In other words, if one forgets about the
microscopic wrinkles of the filament by considering only its
low-resolution shape ${\bf R}(x)$ (the Fourier transform of the
microscopic shape truncated at the cutoff~$\Lambda$), one is actually
considering a system that has the same type of elastic free-energy as
a tense string. Indeed, in the minimum energy configuration, the
low-resolution shape of the filament is straight and its tension is
given by
\begin{equation}
\label{sigma0low}
\sigma_0\simeq\frac{1}{4\beta^2\kappa}\left(\frac{L}{\bar S-L}\right)^2,
\end{equation}
which follows from Eq.\,(\ref{sigmalow}) by setting $S=L$. Again, we
note that this effective tension is independent of the
resolution~$\Lambda$ (as long as $\Lambda\ll\Lambda_c$). For small
deformations about this straight state, one can neglect the fact that
$\sigma$ increases with $S$ as one pulls on the ``string''. Then, the
elastic free-energy of the filament, as seen at low-resolution, is
\begin{equation}
\bar F[{\bf R}]\simeq
\int_0^L\!dx\,\frac{1}{2}\sigma_0\left(\frac{d{\bf R}}{dx}\right)^2.
\end{equation}
In this last expression, we have neglected the remaining energy
associated with the bending of the filament [see
Eq.\,(\ref{FbaravecphideS})]. It is actually negligible in the tense
regime: for all the permitted wavevectors ($q<\Lambda$), we have
$q\ll\Lambda_c$, and therefore $\sigma_0q^2\gg\kappa q^4$ since
$\sqrt{\sigma_0/\kappa}\approx\Lambda_c$, as can be seen from
Eqs.~(\ref{rangevalid}) and~(\ref{sigma0low}).

Although this paper is focused on the concept of effective tension for a
wormlike chain viewed at a low resolution, Eqs.\ (\ref{FbaravecphideS}),
(\ref{phideSexact}) and~(\ref{sigmadeS}) actually give the effective
free-energy fit to describe an attached wormlike chain viewed at
\textit{any} resolution.

It is interesting to compare the present results with those of previous
works. For instance, using the wormlike chain model, \textcite{marko95}
have calculated the force~$f$ needed to extend a DNA molecule attached
at its extremities. In agreement with experiments~\cite{smith92}, they
found that the end-to-end distance approaches the DNA contour-length as
$f^{-1/2}$. Precisely, they obtained
$(4fL_p/k_\mathrm{B}T)^{-1/2}\simeq1-L/\bar S$ in the strong force
limit. Since $\beta\kappa\equiv L_p$, Eq.~(\ref{sigma0low}) reproduces
exactly the same result, upon identifying the effective tension
$\sigma_0$ with the force $f$. 

Since the early work of Helfrich~\cite{helfrich85,peliti85}, a large
literature has been devoted to the renormalization of the elastic
constants of membranes. No similar work has been performed on wormlike
chains, however: this is in part because it is possible to directly
calculate the tangent-tangent correlation function by analogy with the
diffusion on a unit sphere~\cite{harris66}. A large amount of work has
been devoted instead to the interpolation between the wormlike and the
Gaussian chain~\cite{bawendi85,aragon85,ha95}. To the best of my
knowledge, however, there has been no previous analysis in the spirit of
the present paper.

\acknowledgements{I wish to thank A. Ajdari, T. B. Liverpool, T.
Lubensky and L. Peliti for useful discussions.}


\begin{thebibliography}{23}
\expandafter\ifx\csname natexlab\endcsname\relax\def\natexlab#1{#1}\fi
\expandafter\ifx\csname bibnamefont\endcsname\relax
  \def\bibnamefont#1{#1}\fi
\expandafter\ifx\csname bibfnamefont\endcsname\relax
  \def\bibfnamefont#1{#1}\fi
\expandafter\ifx\csname citenamefont\endcsname\relax
  \def\citenamefont#1{#1}\fi
\expandafter\ifx\csname url\endcsname\relax
  \def\url#1{\texttt{#1}}\fi
\expandafter\ifx\csname urlprefix\endcsname\relax\def\urlprefix{URL }\fi
\providecommand{\bibinfo}[2]{#2}
\providecommand{\eprint}[2][]{\url{#2}}

\bibitem[{\citenamefont{Arag{\'o}n and Pecora}(1985)}]{aragon85}
\bibinfo{author}{\bibnamefont{Arag{\'o}n}, \bibfnamefont{S.~R.}}, and
  \bibinfo{author}{\bibfnamefont{R.}~\bibnamefont{Pecora}},
  \bibinfo{year}{1985}, \bibinfo{journal}{Macromolecules}
  \textbf{\bibinfo{volume}{18}}, \bibinfo{pages}{1868}.

\bibitem[{\citenamefont{Bawendi and Freed}(1985)}]{bawendi85}
\bibinfo{author}{\bibnamefont{Bawendi}, \bibfnamefont{M.~G.}}, and
  \bibinfo{author}{\bibfnamefont{K.~F.} \bibnamefont{Freed}},
  \bibinfo{year}{1985}, \bibinfo{journal}{J. Chem. Phys.}
  \textbf{\bibinfo{volume}{83}}, \bibinfo{pages}{2491}.

\bibitem[{\citenamefont{Cai} \emph{et~al.}(1994)\citenamefont{Cai, Lubensky,
  Nelson, and Powers}}]{cai94}
\bibinfo{author}{\bibnamefont{Cai}, \bibfnamefont{W.}},
  \bibinfo{author}{\bibfnamefont{T.~C.} \bibnamefont{Lubensky}},
  \bibinfo{author}{\bibfnamefont{P.}~\bibnamefont{Nelson}}, and
  \bibinfo{author}{\bibfnamefont{T.}~\bibnamefont{Powers}},
  \bibinfo{year}{1994}, \bibinfo{journal}{J. Phys. II France}
  \textbf{\bibinfo{volume}{4}}, \bibinfo{pages}{931}.

\bibitem[{\citenamefont{Chaikin and Lubensky}(1995)}]{chaikin_lubensky_book}
\bibinfo{author}{\bibnamefont{Chaikin}, \bibfnamefont{P.~M.}}, and
  \bibinfo{author}{\bibfnamefont{T.~C.} \bibnamefont{Lubensky}},
  \bibinfo{year}{1995}, \emph{\bibinfo{title}{Principles of Condensed Matter
  Physics}} (\bibinfo{publisher}{Cambridge University Press},
  \bibinfo{address}{Cambridge}).

\bibitem[{\citenamefont{David and Leibler}(1991)}]{david91}
\bibinfo{author}{\bibnamefont{David}, \bibfnamefont{F.}}, and
  \bibinfo{author}{\bibfnamefont{S.}~\bibnamefont{Leibler}},
  \bibinfo{year}{1991}, \bibinfo{journal}{J. Phys. II France}
  \textbf{\bibinfo{volume}{1}}, \bibinfo{pages}{959}.

\bibitem[{\citenamefont{Doi and Edwards}(1989)}]{doi_edwards_book}
\bibinfo{author}{\bibnamefont{Doi}, \bibfnamefont{M.}}, and
  \bibinfo{author}{\bibfnamefont{S.~F.} \bibnamefont{Edwards}},
  \bibinfo{year}{1989}, \emph{\bibinfo{title}{The Theory of Polymer Dynamics}}
  (\bibinfo{publisher}{Clarendon Press}, \bibinfo{address}{Oxford}).

\bibitem[{\citenamefont{Fournier} \emph{et~al.}(2001)\citenamefont{Fournier,
  Ajdari, and Peliti}}]{fournier01}
\bibinfo{author}{\bibnamefont{Fournier}, \bibfnamefont{J.-B.}},
  \bibinfo{author}{\bibfnamefont{A.}~\bibnamefont{Ajdari}}, and
  \bibinfo{author}{\bibfnamefont{L.}~\bibnamefont{Peliti}},
  \bibinfo{year}{2001}, \bibinfo{journal}{Phys. Rev. Lett.}
  \textbf{\bibinfo{volume}{86}}, \bibinfo{pages}{4970}.

\bibitem[{\citenamefont{de~Gennes and Prost}(1993)}]{deGennes_book}
\bibinfo{author}{\bibnamefont{de~Gennes}, \bibfnamefont{P.-G.}}, and
  \bibinfo{author}{\bibfnamefont{J.}~\bibnamefont{Prost}},
  \bibinfo{year}{1993}, \emph{\bibinfo{title}{The Physics of Liquid Crystals}}
  (\bibinfo{publisher}{Clarendon}, \bibinfo{address}{Oxford}).

\bibitem[{\citenamefont{Gittes} \emph{et~al.}(1993)\citenamefont{Gittes,
  Mickey, Nettleton, and Howard}}]{gittes93}
\bibinfo{author}{\bibnamefont{Gittes}, \bibfnamefont{F.}},
  \bibinfo{author}{\bibfnamefont{B.}~\bibnamefont{Mickey}},
  \bibinfo{author}{\bibfnamefont{J.}~\bibnamefont{Nettleton}}, and
  \bibinfo{author}{\bibfnamefont{J.}~\bibnamefont{Howard}},
  \bibinfo{year}{1993}, \bibinfo{journal}{J. Cell. Biol.}
  \textbf{\bibinfo{volume}{120}}, \bibinfo{pages}{923}.

\bibitem[{\citenamefont{Goldenfeld}(1992)}]{goldenfeld_book}
\bibinfo{author}{\bibnamefont{Goldenfeld}, \bibfnamefont{N.}},
  \bibinfo{year}{1992}, \emph{\bibinfo{title}{Lectures on Phase Transitions and
  the Renormalization Group}} (\bibinfo{publisher}{Frontiers in
  Physics---Perseus Books, Reading}, \bibinfo{address}{Massachusetts}).

\bibitem[{\citenamefont{Ha and Thirumalai}(1995)}]{ha95}
\bibinfo{author}{\bibnamefont{Ha}, \bibfnamefont{B.-Y.}}, and
  \bibinfo{author}{\bibfnamefont{D.}~\bibnamefont{Thirumalai}},
  \bibinfo{year}{1995}, \bibinfo{journal}{J. Chem. Phys.}
  \textbf{\bibinfo{volume}{103}}, \bibinfo{pages}{9408}.

\bibitem[{\citenamefont{Hagerman}(1988)}]{hagerman88}
\bibinfo{author}{\bibnamefont{Hagerman}, \bibfnamefont{P.~J.}},
  \bibinfo{year}{1988}, \bibinfo{journal}{Annu. Rev. Biophys. Biophys. Chem.}
  \textbf{\bibinfo{volume}{17}}, \bibinfo{pages}{265}.

\bibitem[{\citenamefont{Harris and Hearst}(1966)}]{harris66}
\bibinfo{author}{\bibnamefont{Harris}, \bibfnamefont{R.~A.}}, and
  \bibinfo{author}{\bibfnamefont{J.~E.} \bibnamefont{Hearst}},
  \bibinfo{year}{1966}, \bibinfo{journal}{J. Chem. Phys.}
  \textbf{\bibinfo{volume}{44}}, \bibinfo{pages}{2595}.

\bibitem[{\citenamefont{Helfrich}(1985)}]{helfrich85}
\bibinfo{author}{\bibnamefont{Helfrich}, \bibfnamefont{W.}},
  \bibinfo{year}{1985}, \bibinfo{journal}{J. Physique}
  \textbf{\bibinfo{volume}{46}}, \bibinfo{pages}{1263}.

\bibitem[{\citenamefont{Israelachvili}(1991)}]{israelachvili_book}
\bibinfo{author}{\bibnamefont{Israelachvili}, \bibfnamefont{J.}},
  \bibinfo{year}{1991}, \emph{\bibinfo{title}{Intermolecular \& Surface
  Forces}} (\bibinfo{publisher}{Academic Press}, \bibinfo{address}{New York}).

\bibitem[{\citenamefont{K{\"a}s} \emph{et~al.}(1994)\citenamefont{K{\"a}s,
  Strey, and Sackmann}}]{kas94}
\bibinfo{author}{\bibnamefont{K{\"a}s}, \bibfnamefont{J.}},
  \bibinfo{author}{\bibfnamefont{H.}~\bibnamefont{Strey}}, and
  \bibinfo{author}{\bibfnamefont{E.}~\bibnamefont{Sackmann}},
  \bibinfo{year}{1994}, \bibinfo{journal}{Nature}
  \textbf{\bibinfo{volume}{368}}, \bibinfo{pages}{226}.

\bibitem[{\citenamefont{Kratky and Porod}(1949)}]{kratky49}
\bibinfo{author}{\bibnamefont{Kratky}, \bibfnamefont{O.}}, and
  \bibinfo{author}{\bibfnamefont{G.}~\bibnamefont{Porod}},
  \bibinfo{year}{1949}, \bibinfo{journal}{Recl. Trav. Chim. Pays-Bas}
  \textbf{\bibinfo{volume}{68}}, \bibinfo{pages}{1106}.

\bibitem[{\citenamefont{Landau and
  Lifshitz}(1971)}]{landau_lifshitz_phystat_book}
\bibinfo{author}{\bibnamefont{Landau}, \bibfnamefont{L.~D.}}, and
  \bibinfo{author}{\bibfnamefont{E.~M.} \bibnamefont{Lifshitz}},
  \bibinfo{year}{1971}, \emph{\bibinfo{title}{Statistical Physics}}
  (\bibinfo{publisher}{Pergamon}, \bibinfo{address}{Oxford}).

\bibitem[{\citenamefont{Marko and Siggia}(1995)}]{marko95}
\bibinfo{author}{\bibnamefont{Marko}, \bibfnamefont{J.~F.}}, and
  \bibinfo{author}{\bibfnamefont{E.~D.} \bibnamefont{Siggia}},
  \bibinfo{year}{1995}, \bibinfo{journal}{Macromolecules}
  \textbf{\bibinfo{volume}{28}}, \bibinfo{pages}{8759}.

\bibitem[{\citenamefont{Peliti and Leibler}(1985)}]{peliti85}
\bibinfo{author}{\bibnamefont{Peliti}, \bibfnamefont{L.}}, and
  \bibinfo{author}{\bibfnamefont{S.}~\bibnamefont{Leibler}},
  \bibinfo{year}{1985}, \bibinfo{journal}{Phys. Rev. Lett.}
  \textbf{\bibinfo{volume}{54}}, \bibinfo{pages}{1690}.

\bibitem[{\citenamefont{Seifert}(1995)}]{seifert95}
\bibinfo{author}{\bibnamefont{Seifert}, \bibfnamefont{U.}},
  \bibinfo{year}{1995}, \bibinfo{journal}{Z. Phys. B}
  \textbf{\bibinfo{volume}{97}}, \bibinfo{pages}{299}.

\bibitem[{\citenamefont{Smith} \emph{et~al.}(1992)\citenamefont{Smith, Finzi,
  and Bustamante}}]{smith92}
\bibinfo{author}{\bibnamefont{Smith}, \bibfnamefont{S.~B.}},
  \bibinfo{author}{\bibfnamefont{L.}~\bibnamefont{Finzi}}, and
  \bibinfo{author}{\bibfnamefont{C.}~\bibnamefont{Bustamante}},
  \bibinfo{year}{1992}, \bibinfo{journal}{Science}
  \textbf{\bibinfo{volume}{258}}, \bibinfo{pages}{1122}.

\bibitem[{\citenamefont{Walker}(1970)}]{walker_book}
\bibinfo{author}{\bibnamefont{Walker}, \bibfnamefont{M.}},
  \bibinfo{year}{1970}, \emph{\bibinfo{title}{Mathematical Methods of Physics}}
  (\bibinfo{publisher}{Addison-Wesley}, \bibinfo{address}{New-York}).

\end{thebibliography}
\end{document}